\documentclass[sigconf]{acmart}

\AtBeginDocument{%
  \providecommand\BibTeX{{%
    \normalfont B\kern-0.5em{\scshape i\kern-0.25em b}\kern-0.8em\TeX}}}

\setcopyright{acmlicensed}
\copyrightyear{2024}
\acmYear{2024}
\acmDOI{XXXXXXX.XXXXXXX}

\acmConference[COMPSCI 294-184 '24]{UC Berkeley EECS}{May '24}{Berkeley, CA}
\acmISBN{978-1-4503-XXXX-X/24/04}

\begin{document}

\title{Code Compass: A Study on the Challenges of Navigating Unfamiliar Codebases}

\author{Ekansh Agrawal}
\affiliation{%
  \country{USA}
  \institution{UC Berkeley EECS}}
\email{agrawalekansh@berkeley.edu}

\author{Omair Alam}
\affiliation{%
  \country{USA}
  \institution{UC Berkeley EECS}}
\email{omair@berkeley.edu}

\author{Chetan Goenka}
\affiliation{%
  \country{USA}
  \institution{UC Berkeley EECS}}
\email{cgoenka@berkeley.edu}

\author{Medha Iyer}
\affiliation{%
  \country{USA}
  \institution{UC Berkeley EECS}}
\email{medhaiyer@berkeley.edu}

\author{Ashish Pandian}
\affiliation{%
  \country{USA}
  \institution{UC Berkeley EECS}}
\email{ashishpandian@berkeley.edu}

\author{Bren Paul}
\affiliation{%
  \country{USA}
  \institution{UC Berkeley EECS}}
\email{bren.m@berkeley.edu}

\author{Isabela Moise}
\affiliation{%
  \country{USA}
  \institution{UC Berkeley EECS}}
\email{isabelamoise@berkeley.edu}


\begin{abstract}
In our research, we investigate the challenges that software engineers face during program comprehension, particularly when debugging unfamiliar codebases. We propose a novel tool, CodeCompass, to address these issues. Our study highlights a significant gap in current tools and methodologies, especially the difficulty developers encounter in effectively utilizing documentation alongside code exploration. CodeCompass tackles these challenges by seamlessly integrating documentation within the IDE, offering context-aware suggestions and visualizations that streamline the debugging process. Our formative study demonstrates how effectively the tool reduces the time developers spend navigating documentation, thereby enhancing code comprehension and task completion rates. Future work will focus on automating the process of annotating codebases, creating sandbox tasks, and providing dynamic support. These innovations could potentially transform software development practices by improving the accessibility and efficiency of program comprehension tools.
\end{abstract}




\maketitle

\section{Motivation}

In our research, we delve into the significant task of program comprehension, where developers explore a codebase to understand its functionality. This activity is not only the initial step in an engineer's workflow when adding to a codebase but also the most time-consuming one, with studies indicating that developers spend an average of 58\% of their time on program comprehension activities \cite{xia2017measuring}. Thus, endeavours to make this process as seamless, effective and productive as possible become as significant as the task itself. 

Several factors contribute to the increased time spent on program comprehension, including inadequate or ambiguous documentation and the need to browse multiple pages to find the desired description of a specific function or test case \cite{xia2017measuring}. Given documentation's pervasiveness in the software community, and the potential it has to enable developers to use APIs and libraries efficiently and contribute to updating software \cite{tan2024detecting}, improving its role struck us as a potential point of investigation. Its prominence was also reflected as we conducted our need-finding studies. We noted that developers struggled as they attempted to debug a new codebase, specifically in matching the code they were exploring to pertinent documentation. This was also evident in our need-finding studies, where we observed developers struggling to align the code they were examining with pertinent documentation. This challenge was apparent in their attempts to use a split-screen to track documentation alongside the code, which they often quickly abandoned, as well as in their expressions of feeling overwhelmed and the fleeting nature of their learning once they navigated away from a file. It was recognized that the existing strategy to explore documentation had room for improvement to enhance code readability as an integral part of code maintenance \cite{aggarwal2002integrated}. We noted that developers frequently toggled between the IDE, documentation, and other online resources, dedicating about 19\% of their programming time to surfing the web for information  \cite{rahman2014towards}. TThis consistent task switching was detrimental, as the longer it took, the harder it was for developers to refocus and reconnect their inquiry to the initial code snippet \cite{kohl2020multitasking}. 

Therefore, we designed CodeCompass, a tool integrated into the IDE to help developers navigate code and browse information more efficiently by optimizing documentation. If successful, this project will enable developers to experience a more streamlined program comprehension process, reducing long lapses in consulting documentation and web searches, repeated inquiries about the same code snippet, and confusion about how to approach the task. The tool would provide developers with an informed entry point into the code if their work is driven by a ticket, attach documentation and valuable demonstrations in line with the code in question to enhance the permanence and application of learning. The ultimate goal is to have this interaction with the tool build developers' intuition for functional calls and enhance their understanding of side effects, allowing them to start coding with more confidence and efficiency.

The study of program comprehension is a vast field, with researchers examining various conditions, such as the type of participant, from computer science students \cite{gupta2022cospex} to field-experts, as well as the constraints of the exploration, by locking IDEs or restricting external access. We are particularly interested in the case study of debugging an unfamiliar codebase within the typical tasks of a software engineer. By allowing free-range permission on the resources and tools available to mimic their typical workflow, we identified a unique combination of factors that led to an unaddressed case study. Thus, no prior studies have adequately answered how software engineers experience debugging obstacles in an unfamiliar codebase using their full resources. After conducting our own study, we found an unaddressed need to match documentation to code, necessitating the development of CodeCompass as a solution. 

This conclusion to develop a new tool also stemmed from the insufficient existing solutions to the problem of participants failing to "match the active code they were working on with the pertinent documentation to enhance understanding." The most common, yet least effective solution known to the participants was a split-screen between the documentation and the IDE. However, this often resulted in abandoned efforts after spending considerable time attempting to understand the code in this manner, rather than achieving successful program comprehension. External to the study, an analysis of previously proposed tools can be categorized into three categories: 1) code/information summary, 2) tool-led information filtration, and 3) internalized resource inquiry. 

For example, AutoComment mines comments from a large programming Q\&A site and generates description comments for similar code segments in open-source projects \cite{wong2013autocomment}. CloCom uses code clone detection techniques to discover similar code segments between a target project and a database of existing code from various repositories before extracting comments to generate a relevant description for the target code \cite{wong2015clocom}. Solutions can also be language-specific, such as automatic comment generator tool that receives a Java signature and body of a method, and identifies the content for the summary and generates natural language text that summarizes the method's overall actions \cite{sridhara2010towards}. However, as observed in our study, simply the provision of code-information does not guarantee greater program comprehension, as it can still lead to a feeling of overwhelm. Not only that, but no clues are given to entry points to begin debugging the code, nor particular files/snippets of interest to complete the task, thereby leaving the developer well-informed but at risk of feeling no more confident in completing the ticket. 

The secondary category of 'tool-led information filtration' does allow for the prioritization of resources to present a hierarchy to developers that enables them to reduce their cognitive load, and focus on selected areas of the code. Mylar employs a degree-of-interest model that dynamically adjusts the visibility and display of project elements in the IDE based on their relevance to the active task. Elements not relevant to the current task are de-emphasized or hidden, reducing clutter and focusing the programmer's attention on pertinent information \cite{kersten2006using}. Furthermore, there exists a program that identifies specific code fragments that represent high-level algorithmic actions within a method. By isolating these fragments, it effectively reduces the amount of detailed code a developer needs to analyze directly. For each identified high-level action, the tool generates a succinct natural language description \cite{sridhara2011automatically}. However, these solutions lack a degree of dynamism and interaction that developers expect to enhance their understanding. As mentioned above, during the need-finding study, beyond documentation, participants spent a significant amount of time on online resources, to query and understand previous examples. This was often followed by re-adjusted search prompts that reflected their evolving understanding, yet the aforementioned solutions provide a singular set of recommendations. Thus after the initial processing, the developer is still left to their own devices should they have more questions, struggle to understand the given description, or wish to check their comprehension.  

Finally, there exist internalized resource inquiry, wherein a search-engine-like tool is built into the IDE itself to avoid toggling/ distractions away from the code. SurfClipse is an Eclipse IDE-based web search solution exploits the APIs provided by Google, Yahoo, Bing and StackOverflow, and captures the, context-relevance, popularity and search engine confidence of each candidate result against the encountered programming problem. It then ranks and presents results from these searches based on relevance, to address detected  errors and exceptions in the project, or based on manual queries from the developer \cite{rahman2014towards}. Seahawk focuses more specifically on StackOverflow, and serves as an Eclipse plugin that formulates queries automatically from the active context in the IDE, presents a ranked and interactive list of results, lets users import code samples in discussions through drag \& drop and link Stack Overflow discussions and source code persistently \cite{ponzanelli2013seahawk}. While these solutions allow for a more interactive engagement to debugging, it still lacks reference to documentation, thereby underutilizing an important source. Furthermore, it remains agnostic to the task assigned to the developer, only addressing the immediate activity/error, with no insight into how to begin or which files/snippets to begin with, nor their functionality. This implies that program comprehension on an unfamiliar codebase while debugging still remains a difficult task, thereby necessitating the CodeCompass tool. 

\section{Related Works}

Our study builds on a substantial body of research that has explored various challenges of program comprehension. The literature reveals a consensus on the significant amount of time and effort developers dedicate to understanding existing codebases, which is crucial for effective software maintenance and development. This background provides a foundation for investigating the specific challenges programmers face and the strategies they employ when navigating new codebases.

Recent research has specifically targeted the difficulties programmers face with unfamiliar codebases \cite{yates2020characterizing}. For instance, a 2017 study provided empirical evidence that developers spend a considerable portion of their time navigating and comprehending new environments, facing specific challenges such as dealing with poorly documented code, understanding complex code structures, and tracing dependencies within the codebase. The literature continually emphasizes the need for more effective tools and practices to enhance codebase exploration and comprehension \cite{ko2006exploratory}. Researchers have explored a variety of software tools designed to aid in program comprehension, including code visualization tools and Integrated Development Environments (IDEs) that offer features like code navigation aids and documentation generators \cite{ma2020domain}. Despite these advancements, a significant gap remains in addressing the full range of challenges that programmers face.

Historically, studies have thoroughly documented the complexity of program comprehension, identifying it as a major bottleneck in software development and maintenance \cite{koenemann1991expert}. Seminal work by Von Mayrhauser and Vans introduced models of program comprehension that describe how developers interact with code to build mental models \cite{von1995program}. This work underscores the importance of understanding the cognitive processes involved in comprehending software systems. A study in 2006 delved into how developers work with unfamiliar code, revealing that they engage in a cyclical process of searching, relating, and collecting information, but often rely on misleading cues leading to failed searches and inefficient navigation. Consequently, developers spend about 35\% of their time navigating the codebase. The study advocates for streamlining the process of finding and managing relevant code, drawing on a new model of program understanding based on information foraging theory \cite{ko2006exploratory}. 

Understanding the cognitive processes involved in code comprehension is essential for developing effective tools and strategies to support developers \cite{neville2003code}. \cite{brooks1982theoretical} proposes a theoretical framework based on knowledge domains and hypothesis generation, where developers form hypotheses about the problem domain and then seek "beacons" or indicators within the code and documentation to verify and refine these hypotheses. This framework highlights the critical role of clear, well-structured, and accessible documentation in guiding the hypothesis verification process and enabling efficient comprehension.

Research has identified prevalent "documentation smells"—such as excessive structural details and fragmented content—that hinder comprehension and contribute to cognitive overload and confusion, impeding developers' ability to effectively work with code. \cite{khan2021automatic} Researchers have identified prevalent "documentation smells" -- such as excessive structural details and fragmented content -- that hinder comprehension and contribute to cognitive overload and confusion, impeding developers' ability to effectively work with code.\cite{uddin2015api}

Studies investigating factors influencing developer productivity further highlight the importance of efficient code comprehension and access to pertinent information \cite{sinha2005incremental} \cite{huang2018salient}. Studies by Robillard \cite{robillard2009makes} \cite{robillard2011field} identify various obstacles encountered by developers when learning new APIs, including challenges related to documentation, API structure, and the technical environment. These studies emphasize the need for comprehensive, well-organized, and easily accessible documentation, aligning with our research goal of providing seamless access to relevant information within the developer's workflow. Similarly, Garousi examines the usage and usefulness of technical documentation in an industrial setting, highlighting the importance of up-to-date, accurate, and precise information.\cite{garousi2015usage} Our research directly addresses this need by providing a tool that dynamically updates and presents relevant documentation based on the specific code elements the developer is exploring.

Furthermore, multiple research tools have stemmed from a study exploring the potential of automatic source code assistance to aid comprehension.\cite{haiduc2010supporting} In addition, further research studies \cite{corritore1999mental} \cite{huber2024leveraging} \cite{bragdon2010code} align with our work by recognizing the value of simplifying the process of code comprehension by automating informative summaries that capture the essence of code elements. Our tool builds upon this concept by debugging with in line documentation and visualizations that facilitate a deeper understanding of the codebase.

A research study by Cheng \cite{cheng2022improves} demonstrates the strong link between code quality and productivity, suggesting that high-quality code, free from technical debt, is easier to understand and navigate. This finding aligns with our research goal of improving code comprehension by providing a tool that helps developers navigate complex codebases and access relevant information efficiently. Additionally, research by Brown \cite{brown2023using} highlights the positive impact of "flow states" on comprehension and problem-solving abilities, emphasizing the need for tools and techniques that promote focus and minimize distractions. Our research directly addresses this need by providing a tool that seamlessly integrates documentation within the IDE, allowing developers to maintain focus and avoid disruptive context switching.

\section{Need Finding Study}

\subsection{User Design Study}

\begin{figure}[h]
\includegraphics[width=8cm]{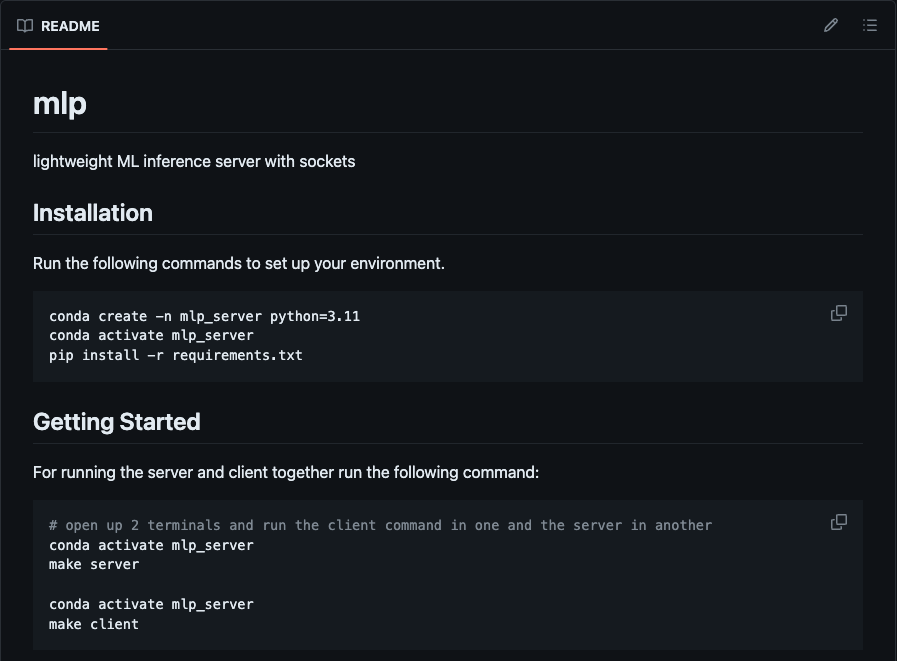}
\caption{Project README.md detailing project installation and setup instructions}
\centering
\label{fig:figure1}
\end{figure}

\begin{figure}[h]
\includegraphics[width=8cm]{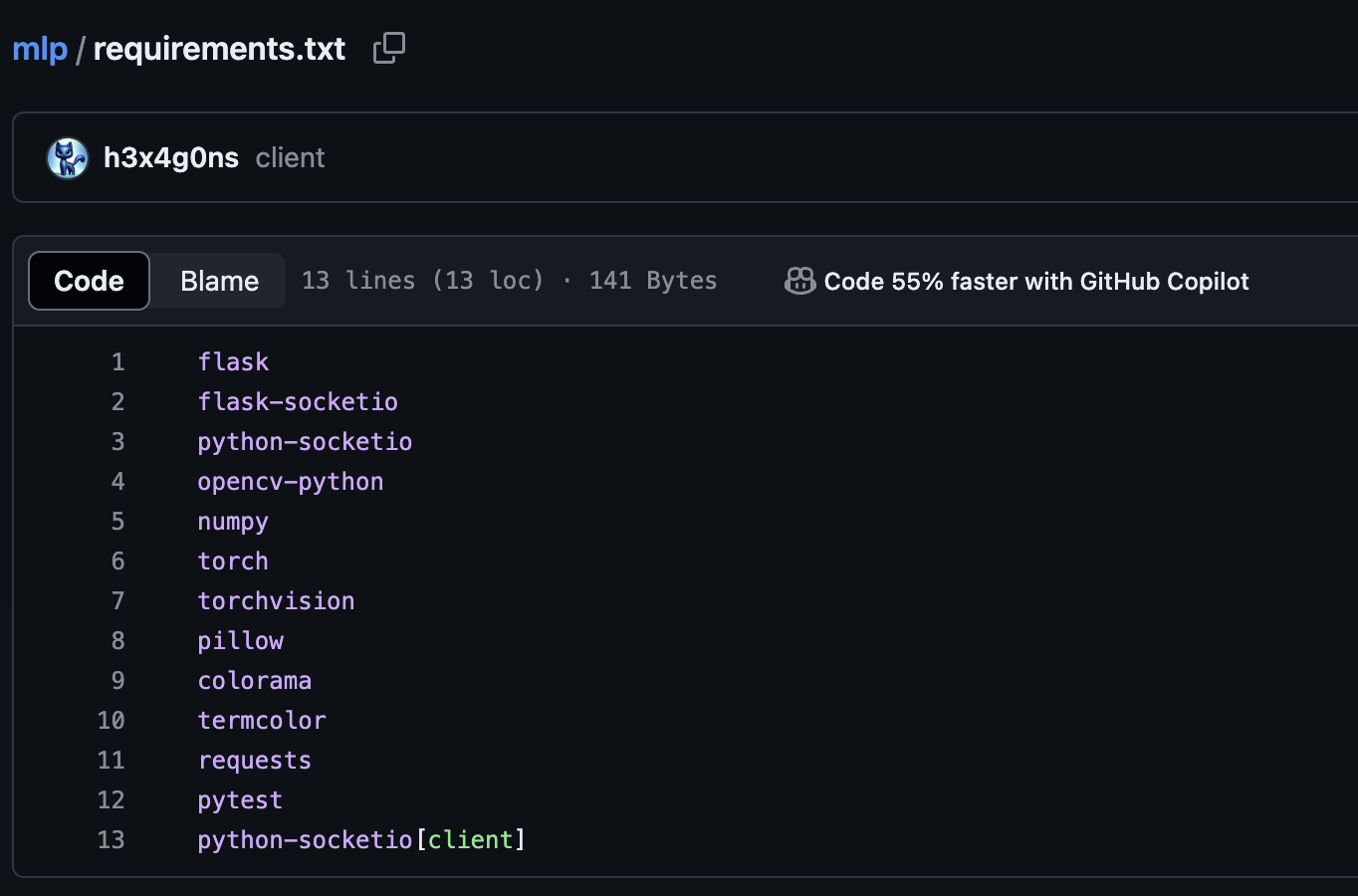}
\caption{Proiect requirements.txt detailing project library depedencies}
\centering
\label{fig:figure2}
\end{figure}

The motivating research question that drove the design of the study was \textit{"What kinds of problems do programmers face when performing bug-fixing maintenance on a previously unseen codebase?"} Our target population thus became intermediate software engineers proficient in Python, all of whom are 18 years old or older. Given the task presented, it is necessary that these participants have had at least one prior encounter/assignment that required them to fix a bug in a codebase, with upwards of 4 years of experience in programming. To ensure that the codebase exploration and debugging could be completed immediately after the start of the study, participants were given installation instructions 24 hours prior to the assigned study time under the Github file \verb|docs/index.md| (Figure \ref{fig:figure1}) that was also echoed in the \verb|README.md|, which included the requirement to download Python, Conda, and the requirements file (Figure \ref{fig:figure2}). 

Data on the target population was collected via evaluation of the potential participants' resume/ development experience, specifically checking for a formal education in Computer Science or software engineering when enquired. Furthermore, special attention was paid to proficiency in Python and codebase exploration familiarity so as to ensure any difficulties they would encounter were not as a result of inexperience. This translated into requiring at least a year of experience working with Python and at least one instance of task maintenance on a new codebase built in Python. 

Using professional and personal connections, the research team recruited 14 engineers for a maximally 60-minutes observation session conducted remotely via Zoom or in-person. For every session, a researcher was assigned to conduct the examination and oversee proceedings. They followed a template script to ensure all steps were followed, attached at the end of the paper as "Script for Study Conduct". Elements included the following steps: 

\begin{figure*}[h]
\includegraphics[width=0.9\textwidth]{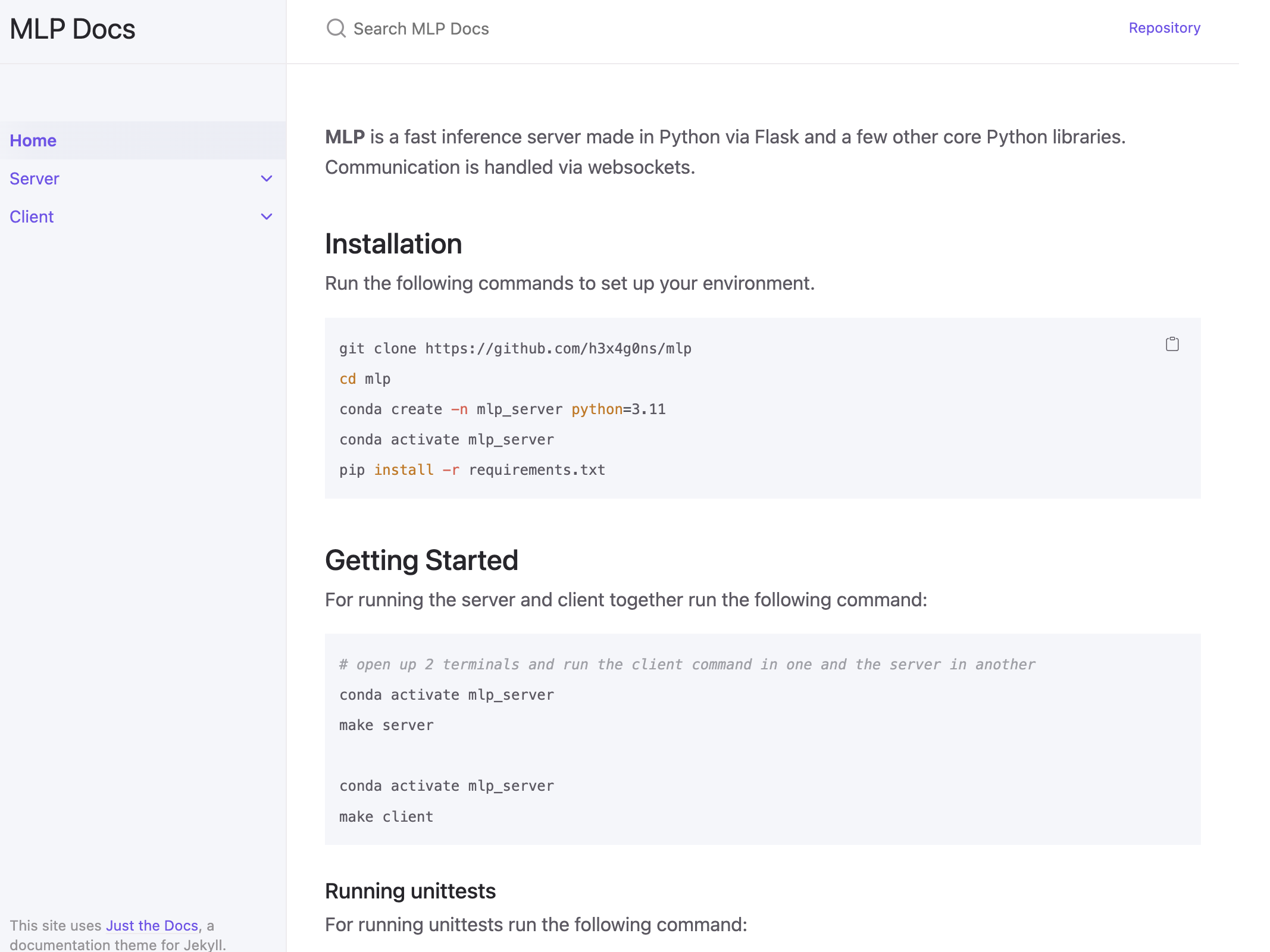}
\caption{Document site for the codebase that was linked to participants}
\centering
\label{fig:figure3}
\end{figure*}

\begin{figure}[h]
\includegraphics[width=8cm]{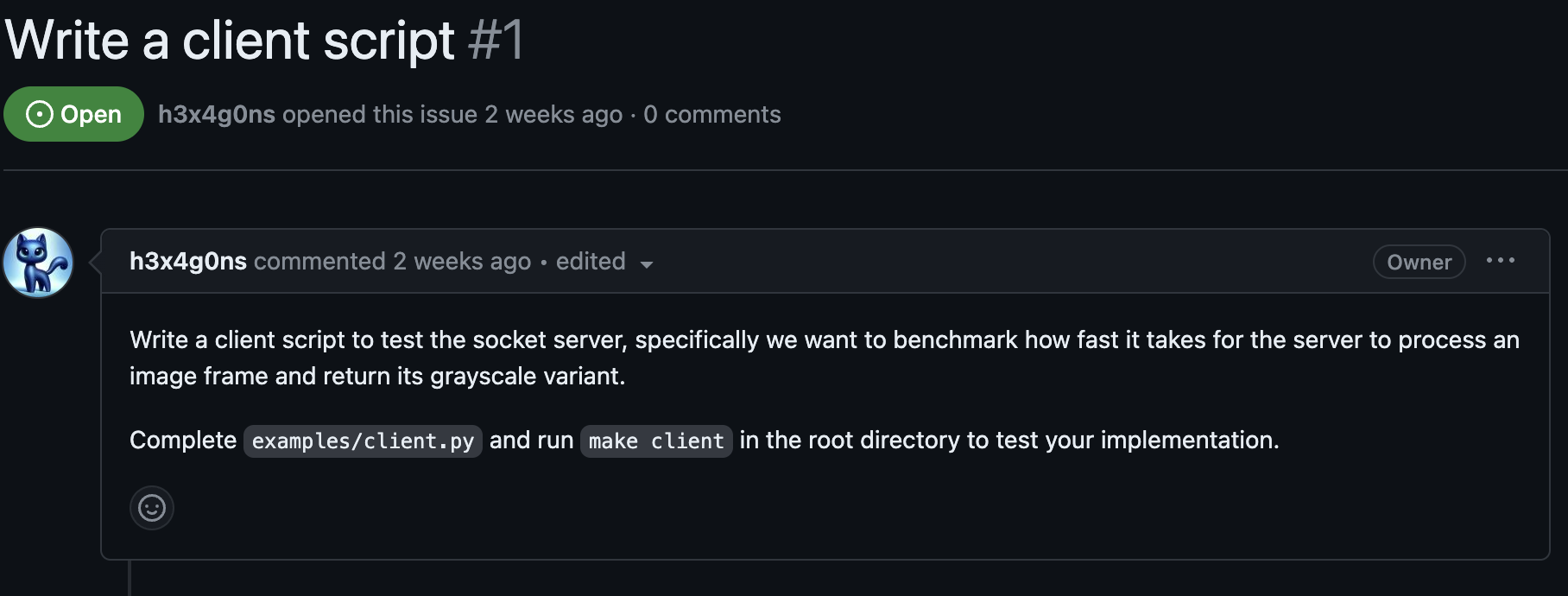}
\caption{GitHub ticket description for Task 1}
\centering
\label{fig:figure4}
\end{figure}

\begin{figure}[h]
\includegraphics[width=8cm]{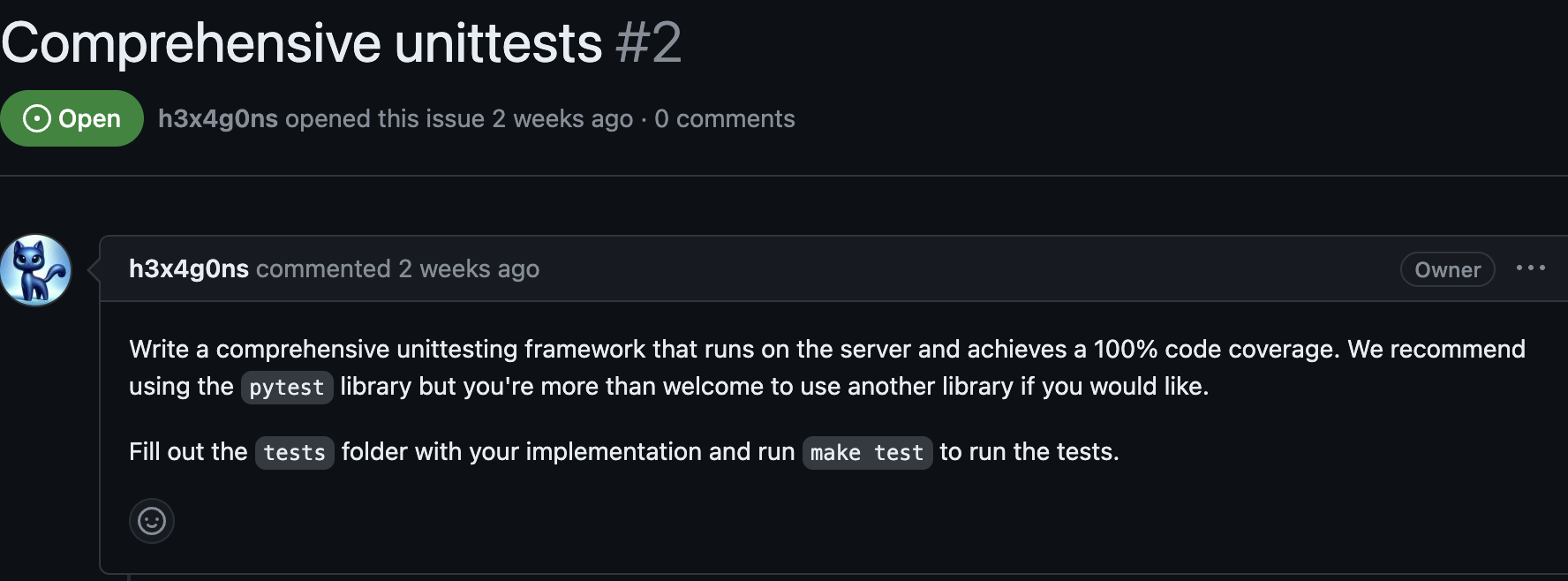}
\caption{GitHub ticket description for Task 2}
\centering
\label{fig:figure5}
\end{figure}

\begin{enumerate}
    \item The study began by introducing the purpose and activity of the study, the mentor-mentee relationship we would be pursuing, alongside confirmation that all the requirements were pre-installed.  
    \item Instructions were then provided, including the request for the participant to begin sharing their screen and narrate their thought processes throughout the session. It was stressed that the participant was empowered to use whatever resources, strategies and tools they typically to their work environment in an attempt to simulate their familiar debugging behaviour.
    \item The participant was then presented with a codebase written in Python that was new and unfamiliar to them \url{https://github.com/h3x4g0ns/mlp}. Alongside access to the code was a visible set of documentation at \url{https://ekanshagrawal.com/mlp/} (Figure \ref{fig:figure3}). Lastly, the participants were provided a link to the GithHub issues for the tasks they would be tackling (\url{https://github.com/h3x4g0ns/mlp/issues}).
    \item The participant was granted permission to begin to navigate the code and complete the two tasks (Figure \ref{fig:figure4} and Figure \ref{fig:figure5}) for at most 40 minutes.
    \item Researchers only interrupted the participant's workflow to clarify their understanding of an action/comments by the latter following its completion. 
    \item The session concluded with a 15-20 minute semi-structured interview with the goal of addressing any remaining questions that the researchers had about the participants' actions or comments. It also served as a time to discuss specific observations from the session, allowing researchers to confirm or refine their interpretations of the participants' actions. 
    \item The culmination of these 14 sessions and their recordings then informed the inductive thematic analysis of the screen and audio recordings via MAXQDA and handwritten analysis.
\end{enumerate}

The motivation behind constructing Task 1 was to provide participants with a relatively quick and actionable fix to bolster confidence and pique curiosity in the functionality of the code. Task 2 was set up in an attempt to enforce and verify comprehensive understanding of the codebase's functionality, to ensure that the research question was appropriately satisfied by having participants thoroughly explore the codebase, and in turn, encounter as many obstacles as possible in debugging. 

To further explain the method of data collection, the screen recording of the participants' computer and their auditory contributions were recorded during the session for future reflexive analysis. This included the .mp4 files for the simultaneous recordings of the participants' audible vocalization of their thoughts during the exploratory session as well as their answers to the interviewers' questions. Concurrently, the entirety of the users' screen was recorded to reflect in real-time the materials the participant was manipulating and reading. For safety and protection purposes, given audio and screen recordings are classified as Protection Level P2, the necessary data usage precautions were followed; that is all files were stored on a shared Berkeley-provided Google Drive which requires organization credential sign-in to access. Furthermore, during reflexive analysis on MAXQDA, local copies on password-protected laptops were saved. However, following the conclusion of thematic coding, these files were deleted from local memory.

Ultimately, a collection of audio and video recordings belonging to 14 participants was gathered, with an average length of 50 minutes per file. During reflexive analysis, researchers paid special attention to the actions performed by each participant as they navigated the new codebase, chronicling every step the user made that was deemed as relevant to the research question, and any vocalized or demonstrated obstacle encountered during debugging.

\subsection{Study Results}

\subsubsection{External Web Usage}

\textbf{Participant struggled with providing sufficient context to assistant LLMs in an attempt to receive conceptual education in a question \& answer format.} During the sessions, participants often turned to LLMs with the purpose of building a stronger conceptual understanding and how to approach the tasks. P6 asked ChatGPT how to use OpenCV and got back a response detailing how to install it along with general applications of OpenCV. P6 decided to pivot to an alternate method of understanding OpenCV and moved back to reading the provided documentation. 

However, others like P4, P5, P7, P8, and P11 underwent multiple prompt attempts to craft a question that addressed their specific gap in knowledge. P8 inquired about the functionality of the 'socket.io decorator', and realized their approach was invalid based on LLM's response. Similarly, P4 ventured a multi-pronged question about how to 'calculate the time it takes for the client to send a request to the server and for the server to finish processing and send a response back'. P11 did not know how to get their code to 'listen for responses', thus querying the LLM for "how to set up a python client that listens for a grayscale response event". This common thread of "how-to" question structure yielded more useful answers, as deemed by the participant's expressed satisfaction with the answer as well as the comparatively more advanced degree of progress they could achieve. However, this success also required a level of knowledge/experience to properly set up the prompt. For example, P11 only went to LLMs once they had developed confidence about what they needed and how to outline their request by spending the majority of the time exploring the codebase.

Without this deeper context, participants like P7 and P10 resorted to pasting a combination of dependencies, task workflow, tickets, and starter code into LLMs in an attempt to outsource the tickets. Thus a lack of prompting expertise on the part of the participants was showcased, possibly indicating a lack of deep conceptual understanding. This confusion and surface-level comprehension was perpetuated even after the LLM provided a response, as it was found that some of the code suggestions were incorrect or overtly complex compared to the researchers' solution, mounting further confusion onto the participants.

\textbf{Participants attempted to find examples of implementation of sockets and client-server code.} However, these participants were often met with incomplete or irrelevant material. During the problem-solving process, several participants attempted to search the web as well as prompt LLMs for implementation examples. A popular example was the search for the socket's client- and server-side demonstrations. P1, P6, P11, and P13 all searched for variations of code samples using the socket.io framework. They expressed similar sentiments of frustration and dissatisfaction at the quality and depth of examples. P1 noticed that to understand many of the examples, they "had to already have some level of background knowledge". P11 noted that they saw far more "examples for JavaScript use cases than Python", providing little to no new insight into how to approach the problem. Furthermore, P13 complained that "this API is really bad. There's no examples in it, which is really confusing me." This need was further illustrated by participants such as P11 and P13, who both discussed their desire for end-to-end examples that were more applicable to the tasks at hand, ideally placed within the given documentation and source code repository. 

The other main area participants queried for online assistance was regarding implementation details for specific functions relevant to accomplishing the assigned  task. This was demonstrated in efforts by P7 and P11, both of whom used a web browser for examples of the "emit" and "on" functionality. Similarly, P8 asked the LLM about how to properly apply the \verb|pytest| framework to test the grayscaling feature, yet despite the resulting examples, were unable to adapt the code to their testing needs. In all such circumstances, these observations demonstrated the inefficacy of the participant's attempt to reference external examples to enhance their understanding.

\subsubsection{Code Comprehension}

\textbf{Participants failed to match the active code they were working on with the pertinent documentation to enhance understanding.} Many of the participants struggled with effectively utilizing the documentation provided alongside the codebase when attempting to take advantage of the information in order to gain a deeper understanding of the project structure. For example, P12 spent the majority of their allotted time reviewing the official Socket.IO documentation but expressed their struggle with connecting the information to the client file they were working on. The inability to trace the documentation to the relevant parts of the code, and vice versa, was a recurring issue expressed as 42\% of the participants attempted a split screen approach, thereby hosting 2 windows in order to view the documentation and code simultaneously. However, in all instances, this approach was abandoned after a maximum of 10 minutes, with P3 citing that the split screen process to manually transfer over documentation as inline comments onto the Python file was "taking too long and isn't helping".

The most common documentation strategies observed by the participants when matching documentation to the code were skimming and linear reading. P2 initially expressed enthusiasm for the vast amount of documentation available, but quickly abandoned the attempt to digest it all in a systematic approach, stating it was "too much reading and I don't understand the code any better". Similarly, later on in the study, as P2 was expanding their exploration of the code files, they lamented about forgetting what they had previously read in the documentation for a file they were revisiting. In a shared experience, while P9 was struggling to understand the code, they also turned to the documentation, acknowledging that they "should've gone here [the documentation site] first". Furthermore, P5 dedicated a quarter of their time, which was greater than any other participant, purely reading documentation without referring to code. And yet they expressed little to no satisfaction with the degree/progress of understanding they had acquired when turning to the codebase and applying their new knowledge. These observations from our study demonstrate the participant's struggle to effectively integrate documentation into their code comprehension process

\subsubsection{Enhancing Domain Specific Knowledge}

\textbf{Participants acquired online media tutorials/lessons to improve their understanding of the client socket model.} In the first task of the study, which involved working with the Flask-SocketIO package, several participants struggled to understand the API's usage, especially those with no prior experience with it. P6 began by exploring the Socket.IO documentation, only to switch focus to the grayscale aspect of the task. Meanwhile, P13 delved even deeper into the documentation, yet struggled to grasp the application of methods. In contrast, P7 consulted ChatGPT for insights on how Socket.IO operated across both the client and server sides but continued to encounter errors after implementing the recommendations from the LLM. P14 voiced their frustration, saying, "I understand how the grayscale function works and I know I need to call it in the client, but I'm not too sure how to go about doing that."

The difficulty lay in connecting the various components of Flask-SocketIO, especially without any prior exposure to it. Both P6 and P13 mentioned that in a scenario like this, they would take a short tangent to learn about the details of Socket.IO. In P13's words, "This is the scenario where I would watch YouTube videos to understand sockets again and then maybe come back to this from a high-level understanding of it." P13 and P8 noted that in such situations, they often preferred turning to videos and articles, which acted as virtual teachers and helped them build a conceptual understanding before diving into the code, an approach they sometimes found more beneficial than navigating through API documentation. This need was further reinforced by P11, who found diagrams depicting the control flow and stated that they were more helpful than the documentation. These observations display the trouble software engineers face when interacting with unseen packages or concepts.

\section{Code Compass Design}

\subsection{Tool Design}

CodeCompass is a VSCode extension that provides a framework to connect documentation, the debugging ticket description and the relevant functions/code snippets of the project. It can be accessed on Github under “Acumane/code-compass” or at the following link: \href{https://github.com/Acumane/code-compass.git}{https://github.com/Acumane/code-compass.git}.

It works to serve as a key onboarding tool for developers who are intent on exploring a new codebase with prioritized file selections and a structured walkthrough of the repository relevant to the supplied ticket description. It also works for specific program comprehension activities within a file, wherein developers can select a code snippet and request information about its functionality with both an overall summary of the section, as well as line-by-line analysis. By being able to select the degree of nuance they would like to explore, the developer can track relevant variables and transformations at every step in the program, or be satisfied with a cursory summarized understanding of its general operation. The user will have the option to step into any nested functions and experience the same walkthrough, or step over and simply learn of its output. After going through the entire functionality, the user is brought back to the top and provided with a task to return a specific output. The interactive example exists in an isolated state, with curated inputs that satisfy the parameters up until this function is supposed to be run, with the rest of the code serving as a 'blackbox'. The developer is able to manipulate the function in question and run only this snippet to gauge success, receiving an error or success message. Once this has been completed, the developer can exit out of the tutorial/debugger mode, the edited function reverts to its original form prior to the tool usage, and they can apply their new learning to the codebase. 

Presently, the documentation, mapped descriptions to lines, and relevant tasks to apply one's understanding are all manually written by engineers familiar with the existing codebase. This ensures that high quality information will be passed through to any new user, and functions efficiently during our initial stages of testing the tool. With further progress, the secondary frontier for this tool would include dynamic and multi-generated function/line explanations and examples, that would allow for context-sensitive/informed tutorials without the need for documentation to be fed into the framework. Instead, the codebase and the task ticket is provided to a secondary program, giving it the ability to prioritize and understand the relevant code elements required to satisfy the request. The line-by-line and isolated environments in which the developer can manipulate the function remains the same, however the tutorial can receive feedback, regenerating the wording for line explanations and examples if the user is dissatisfied/remains confused. 

The final addition to maximize the frontier of the tool would be codebase orientation, with a more extended frontend which reads through the codebase and identified objective (either completing a debugging ticket or general program comprehension), in order to generate a prioritized curricula through which the developer can explore within the codebase to learn what is relevant to them to complete their task. This would be provided through identified subsections of the codebase that would require understanding/modification to complete the task, context-dependent lectures/summaries, and internal evaluations of the information/task provided to the user.

\subsection{User Flow}

An illustrative user story harkens back to the need-finding study, specifically right after reading the issue ticket on Github that requests developers to: 

\textit{“Write a client script to test the socket server, specifically we want to benchmark how fast it takes for the server to process an image frame and return its grayscale variant. Complete} \verb|examples/client.py| \textit{and run make client in the root directory to test your implementation.”}

The task they would then attempt to complete would be exploring the aforementioned \verb|examples/client.py| file and understand how it currently works. This often would lead them to open the file, and compare it against the provided documentation. However, the majority of cases led to participants becoming lost as they tracked the various function calls, searched up unknown libraries, and ultimately spent time on irrelevant queries/resources with little progress on comprehension or completing the task. 

\begin{figure}[h]
\includegraphics[width=8cm]{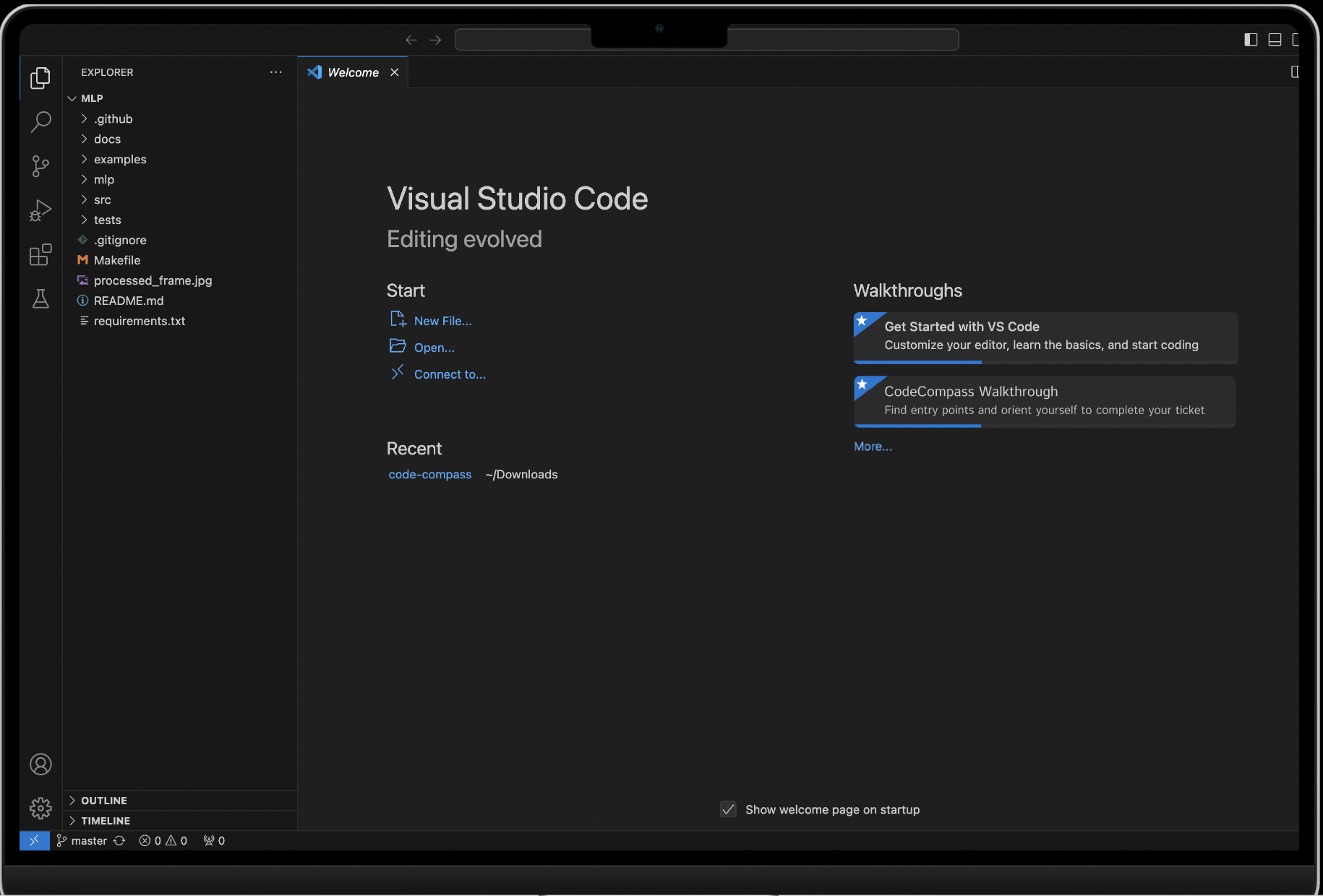}
\caption{Splash screen when loading VSCode with the CodeCompass extension enabled}
\centering
\label{fig:figure6}
\end{figure}

To streamline this process to allow for both greater comprehension and code modification in a relatively shorter amount of time, the user would use CodeCompass as the solution. The following walkthrough showcases the tool's capabilities assuming it has been fully developed and is advanced enough to incorporate LLM input and a codebase orientation capability at its maximal frontier. After reading the ticket, developers would navigate to the VSCode IDE and feed the ticket link into the CodeCompass activation button underneath the 'Walkthroughs' section of VSCode (as shown in Figure \ref{fig:figure6}). 

\begin{figure*}[h]
\includegraphics[width=0.9\textwidth]{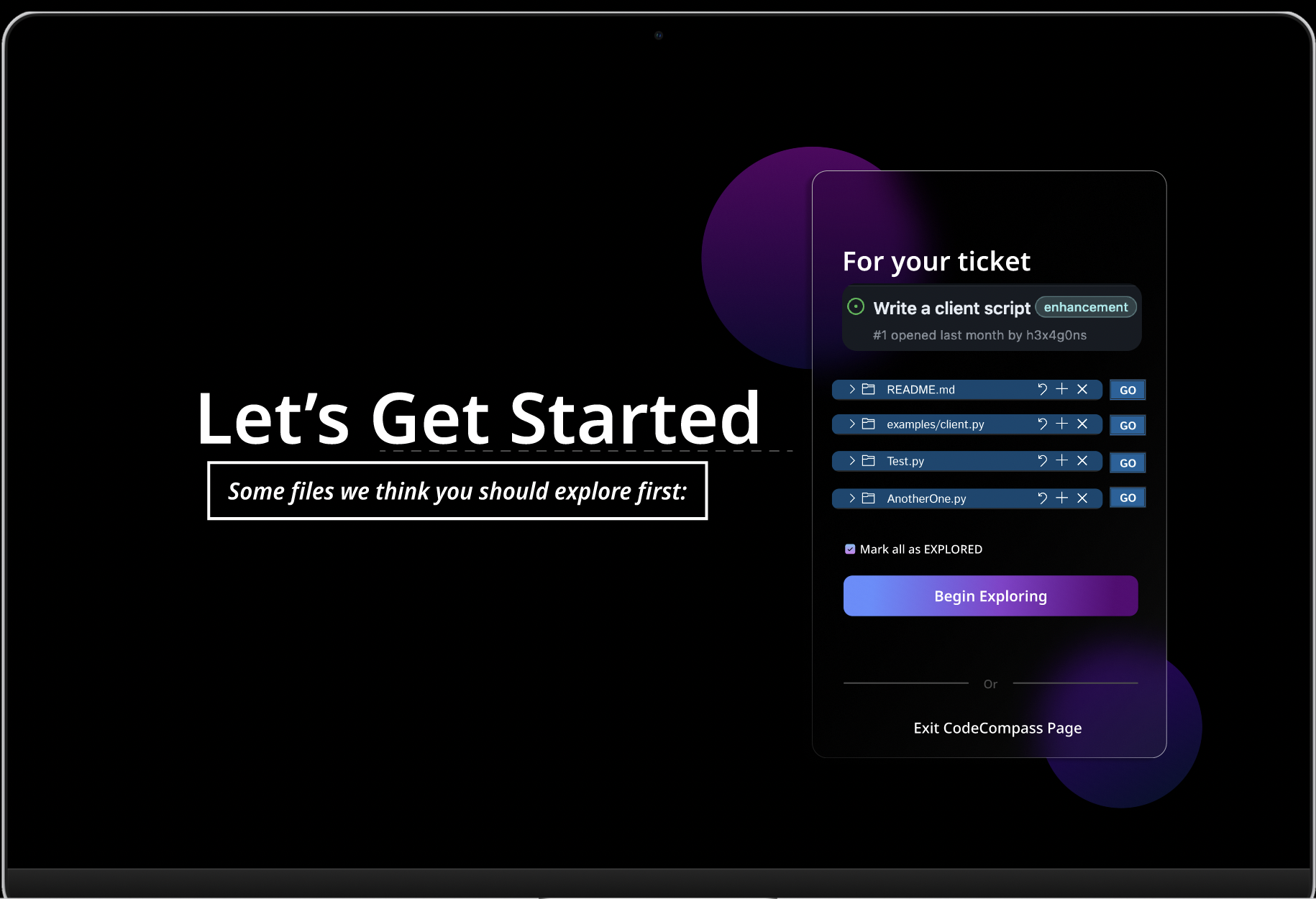}
\caption{Enabling guided walk through once a codebase is loaded}
\centering
\label{fig:figure7}
\end{figure*}

The user will then be greeted by a 'Welcome' screen to CodeCompass' main page within the IDE, with a selection of files to begin exploring either individually by the developer or through a guided walkthrough based on the tool's analysis of the documentation and relevance to the ticket displayed at the top of the right hand side box (as shown in Figure \ref{fig:figure7}). Given this task has the developer intent on exploring \verb|examples/client.py| first, the user can select the 'Go' button displayed next to the file to be directed to the file. Should they have wanted the systematic orientation throughout the codebase with resources relevant to the ticket, they could have instead selected the “Begin Exploring” button, and if they decided to abandon the endeavor altogether, they are also able to 'Exit CodeCompass Page' with the bottom right button.  

\begin{figure*}[h]
\includegraphics[width=0.9\textwidth]{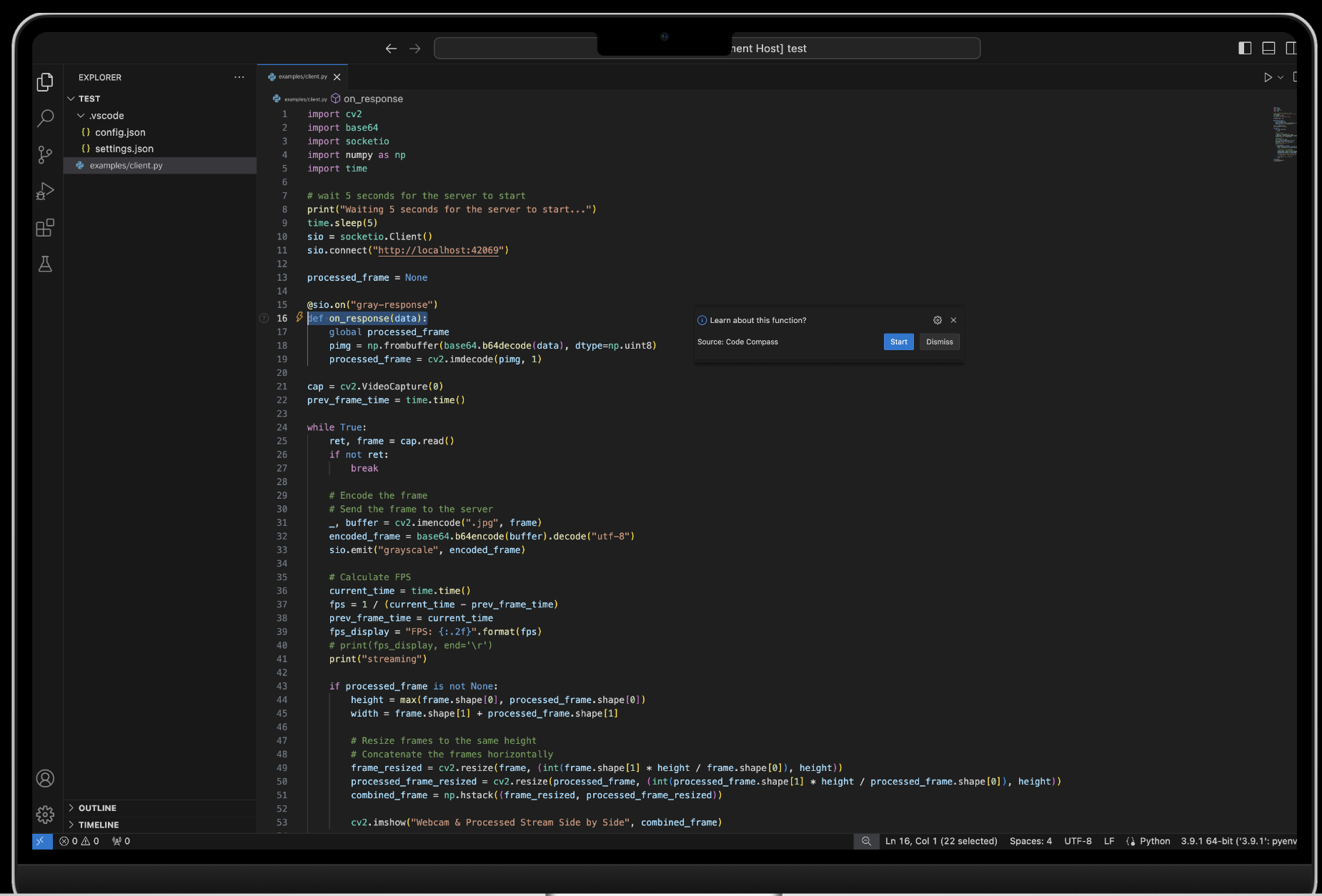}
\caption{Pop-up for activation tutorial once hovering over a function}
\centering
\label{fig:figure8}
\end{figure*}

\begin{figure*}[h]
\includegraphics[width=0.9\textwidth]{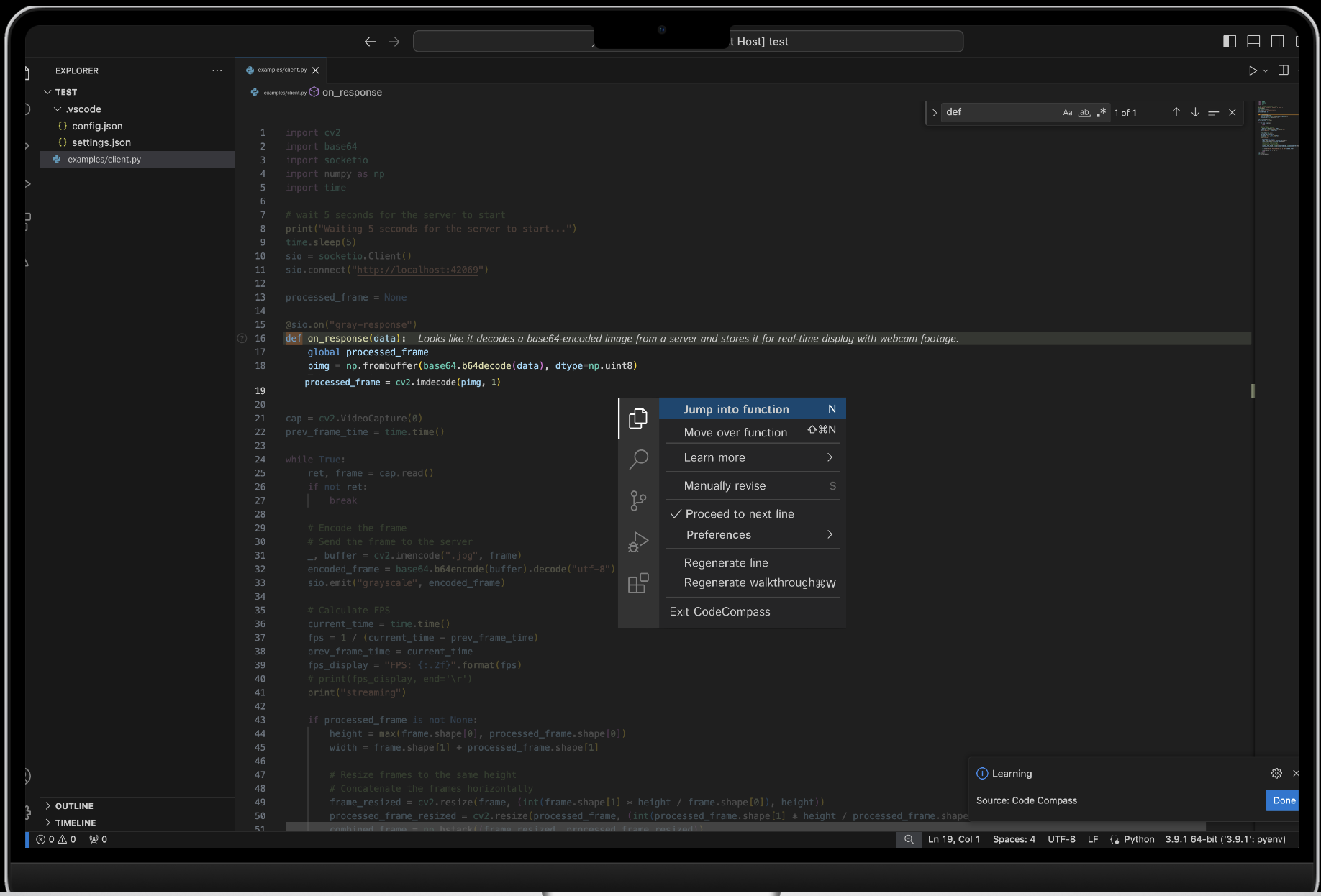}
\caption{Menu option for activating the code lens features and showing line by line description}
\centering
\label{fig:figure9}
\end{figure*}

\begin{figure*}[h]
\includegraphics[width=0.9\textwidth]{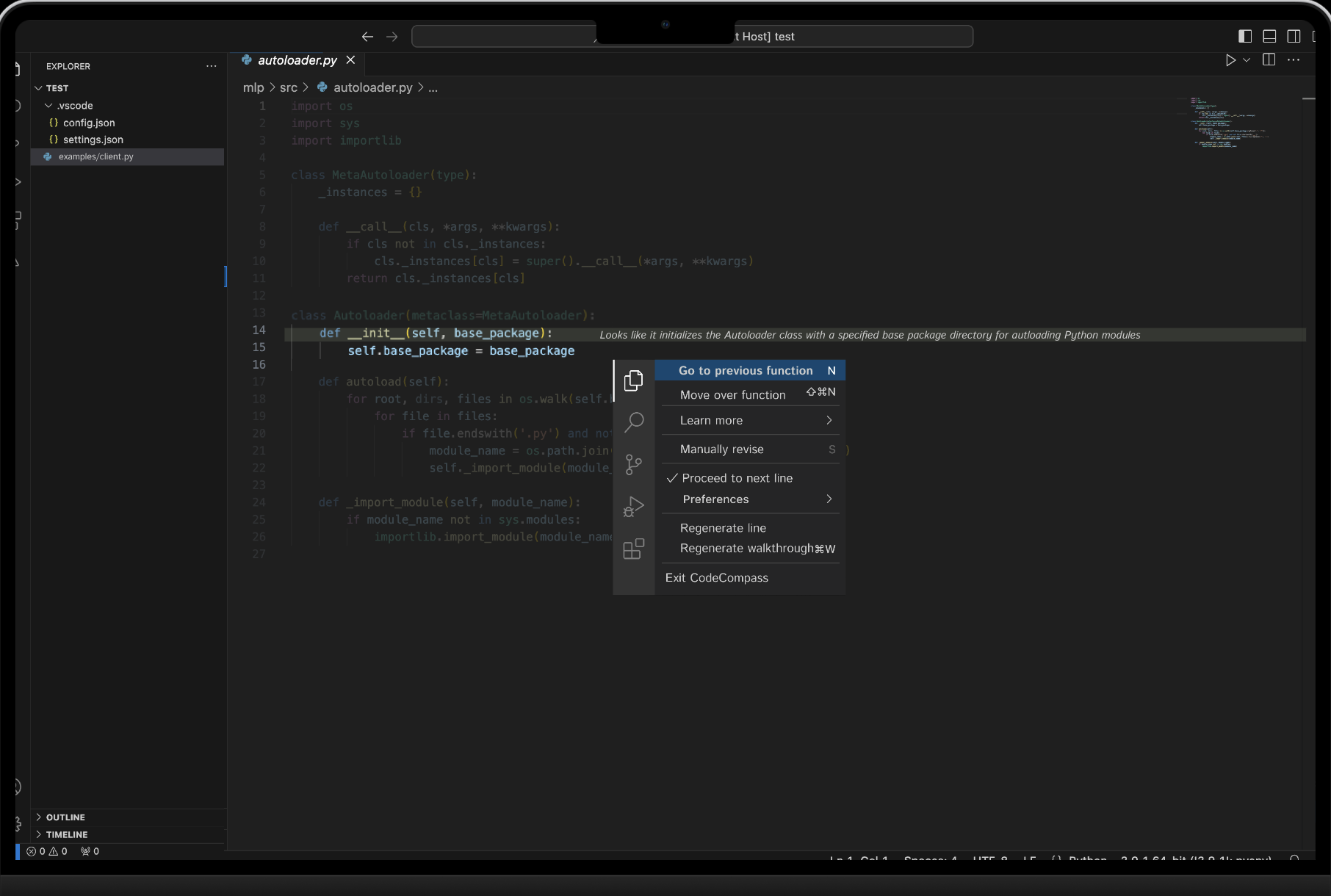}
\caption{Jumping in and out of functions to understand flow}
\centering
\label{fig:figure10}
\end{figure*}

\begin{figure*}[h]
\includegraphics[width=0.9\textwidth]{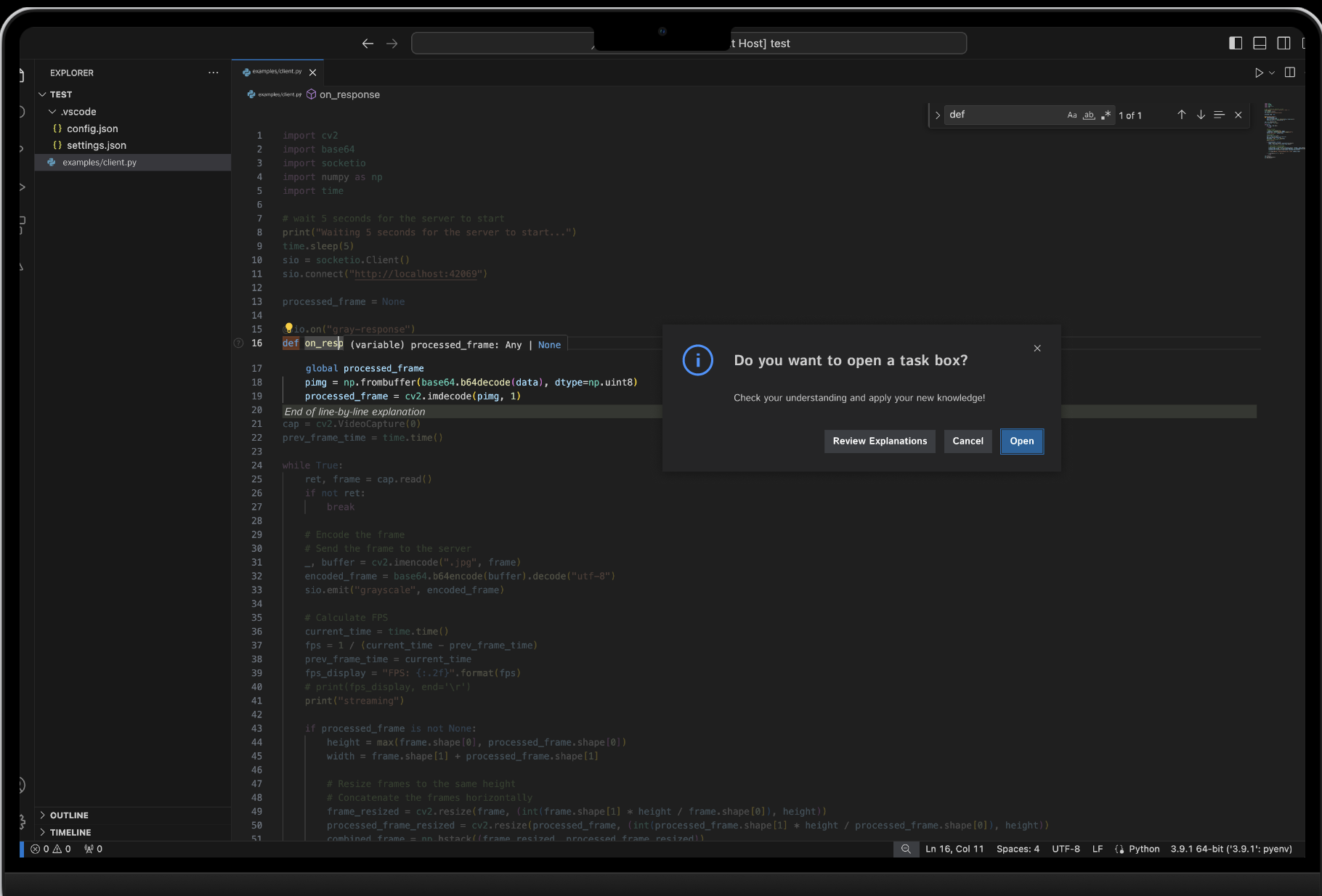}
\caption{Prompt for opening playground}
\centering
\label{fig:figure11}
\end{figure*}

After landing on the \verb|examples/client.py| file, the user can then select any function or code snippet of interest, and trigger a popup notification to 'Learn about this function' (as shown in Figure 8). This changes the screen to darken the rest of the file beyond the snippet of interest to encourage focus, highlighting an initial summary of the entire grouping. Should another function outside of the current file be referenced, the developer has the option to 'Jump into function' to be redirected to the inner function and receive a similar line-by-line analysis (as shown in Figure \ref{fig:figure9}). Other options within the same menu box includes:

\begin{enumerate}
    \item Moving over the function to avoid any further analysis and continue with the code snippet of interest,
    \item Manually revise the line/summary analysis,
    \item Proceeding to the next line,
    \item Requesting a re-worded summary, line explanation or entire walkthrough to be re-generated by the attached LLM
    \item Exiting the tool
\end{enumerate}

\begin{figure*}[h]
\includegraphics[width=0.9\textwidth]{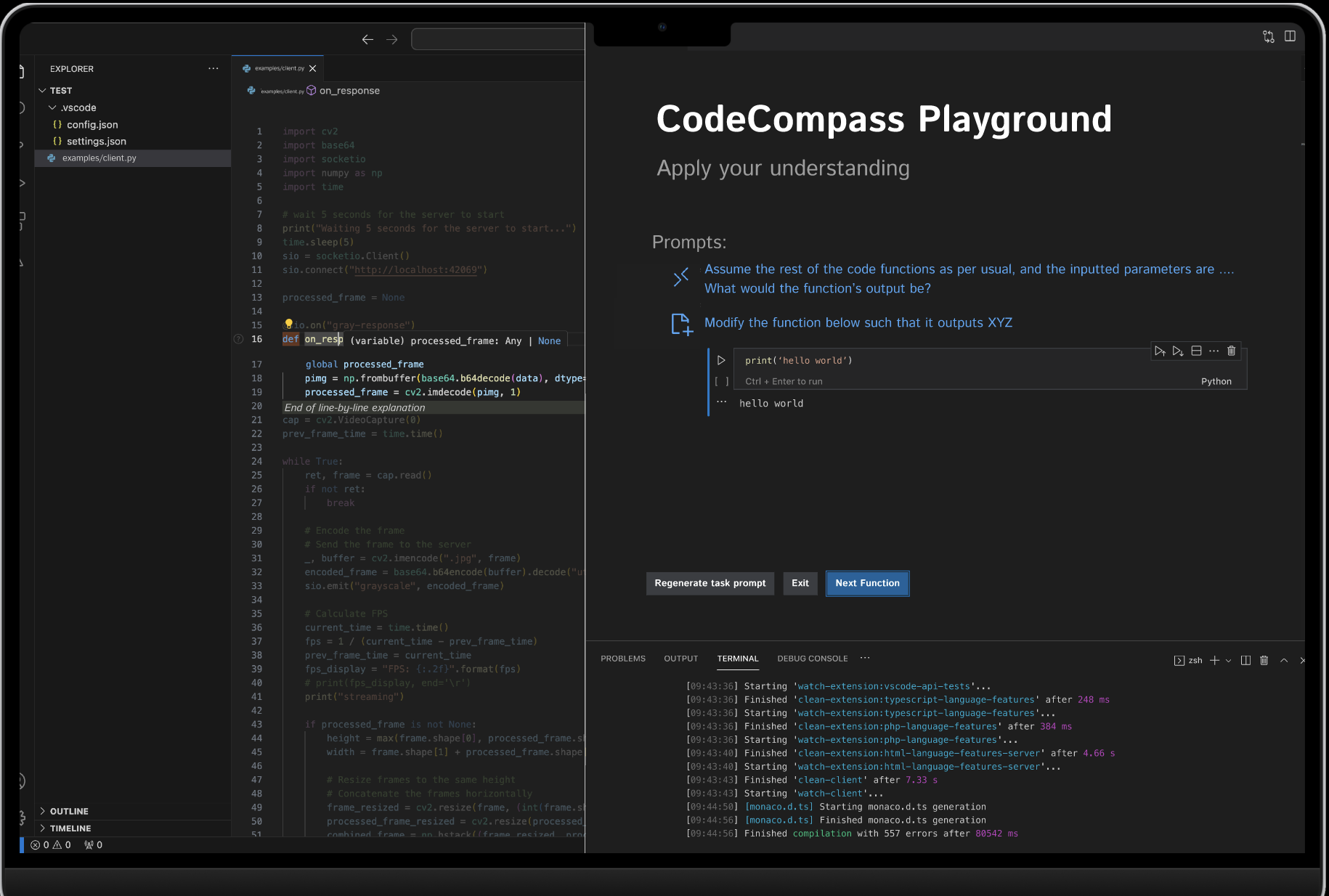}
\caption{Playground to test inputs and outputs against the function}
\centering
\label{fig:figure12}
\end{figure*}

After exploring the inner function, there is an option to return to the initial code snippet/function of interest (as shown in Figure \ref{fig:figure10}). After landing on the last line of analysis, another notification presents itself, asking the developer if they 'want to open a task box' in order to allow them to apply their understanding (as shown in Figure \ref{fig:figure11}). Opening the box results in a 'CodeCompass Playground' asserting itself on the right hand side, alongside a series of prompts that test the developer's understanding (as shown in Figure \ref{fig:figure12}). This can include a 'fill in the blank' assignment, or a mini code snippet available for modification to test the user's understanding. Ultimately, they receive feedback on their submitted answer through the terminal screen at the bottom of the page, or a notification message within the page itself. At any time, the user is also able to: 

\begin{enumerate}
    \item Request new/reworded task prompts, 
    \item Exit the task box
    \item Move to the next function for analysis
\end{enumerate}

Thus the task of gaining an understanding of \verb|examples/client.py| has been thoroughly explored and tailored to the developer's degree of interest to enhance program comprehension. 

To provide an illustration of how the tool at its current state would handle the same use case, the initialization of the tool is the same, as the developers explore their new codebase, and select a particular code snippet. A popup on the right hand side will open a prompt to 'learn about this function' (Figure 8). Selecting the 'Start' button begins the walkthrough, with the relevant description previously fed into the tool by a codebase expert placed alongside the relevant line of code (Figure 9). The user is able to use the 'Continue' button right above the highlighted line of code to traverse through the lines, or the 'exit' button alongside it to quit the experience. The application piece then appears with a task also crafted and fed into the tool by an expert on the codebase. To gauge progress, the developer clicks the VSCode run icon, and receives an assertion error until they have successfully solved the task. 

\subsection{Design Decisions}

Designing a tool like CodeCompass involves numerous key decisions that intertwine technological capabilities with educational strategies, ensuring that the tool is both effective and intuitive for developers. 

When considering the pedagogical framework to support learning and comprehension within the isolated task section of the experience, we drew inspiration from well-known educational platforms that teach programming such as Khan Academy as well as teaching best practices. When building the active learning environment, we wanted to include immediate feedback, such that developers are quickly provided with information on their performance, which helps to reinforce learning or correct misunderstandings. Thus, there is an instant check within the task box to confirm the users' progress with an error or 'success' message after running the modified code in the sandbox. The effect is, to significantly enhance the static descriptions provided in the initial summaries. This will ideally improve developer understanding and retention of information \cite{prince2004does}. 

Similarly, the combination of both the initial snippet summary of the functionality and/or the line-by-line walkthrough of the code alongside the task box had us design for incremental learning and scaffolding. In other words, helping developers increase their program comprehension and skills by providing temporary resources/aides that are gradually removed as competence grows. This manifested in creating a task box that provides several types of assignments the developer can try, be it 'fill in the blank' or active code modification. Following the completion of the task, the user is assumed to have this new information retained, and so the summaries, analysis and box are all removed and any modification reverts back. Thus the developer gradually has support structures introduced and then removed to challenge and cement their understanding \cite{wood1976role}. 

Another considered design element was integration and user interface design. Specifically, designing CodeCompass to ensure it is helpful without being obtrusive, integrating seamlessly into developers' workflows. This was done by applying design principles and theories. To make the tool intuitive, the design incorporates clear affordances by using familiar interaction models from VSCode, ensuring that developers can easily understand what actions are possible. Signifiers are used to indicate where actions should take place, such as icons or buttons specific to debugging tasks, helping guide user behavior without confusion. The tool's features are mapped in a way that aligns with the developers' expectations based on their experience with VSCode, such as using similar commands to toggle views or execute functions. Feedback is immediate and informative, crucial for interactive elements like live code editing or debugging, providing users with clear and immediate information about the effects of their actions. Following principles on cognitive load, the text within the tool is kept minimal and focused, aiding in reducing distractions and enhancing concentration during critical tasks \cite{norman2013design}. Finally, within the study, we recognized the diverse workflow speeds among developers. Thus, the tool includes  an always-visible exit button allows them to quickly disengage from its functions, a decision supported by the principle of user control in interaction design.

The third design consideration was design consistency and usability, such that CodeCompass' aesthetic within VSCode provided a frictionless user experience. By maintaining the visual and operational consistency with VSCode, CodeCompass minimizes the learning curve for new users. This consistency in design ensures that once a developer learns one part of the environment, the same patterns apply to learning other parts, significantly easing the user journey. Adhering to established UI standards of VSCode, such as color schemes, typography, and iconography, the tool aligns with the developers' pre-existing knowledge and expectations. This adherence not only enhances usability but also fosters an intuitive interaction environment where developers can focus more on learning and debugging rather than navigating the tool. Furthermore, CodeCompass' design utilizes simple and familiar symbols, which serve as effective signifiers without overwhelming the user. This choice supports the usability principle of reducing cognitive load while maintaining functionality.

Ultimately, the design of CodeCompass s meticulously designed in response to explicit challenges uncovered in a detailed need-finding study that revealed developers often struggle with the disconnection between documentation and active code. The tool informatively integrates debugging ticket descriptions with relevant code snippets and documentation, directly addressing the study's finding where 42\% of participants tried, but quickly abandoned, a split-screen approach as inefficient and unhelpful. CodeCompass not only offers structured walkthroughs of repositories tailored to the developers' current tasks but also provides a unique interactive feature that lets developers choose the depth of detail—from a broad overview to an in-depth, line-by-line analysis. This directly counters the frustration expressed by participants who felt overwhelmed by documentation, as one remarked, it was "too much reading and I don't understand the code any better."

Furthermore, CodeCompass introduces an isolated testing environment that allows for safe manipulation and testing of code snippets. This functionality responds to the need for applicable, hands-on examples cited in the study, where developers like P13 expressed dissatisfaction with available API examples, saying, "There's no examples in it, which is really confusing me." By enabling users to experiment with changes in a controlled setting where modifications can be reverted, the tool ensures that learning is both practical and risk-free. This is crucial since many participants indicated a lack of confidence in applying newly learned concepts directly to their tasks.

\section{Implementation}

CodeCompass is a VSCode extension written in TypeScript. When the extension is loaded, the \verb|activate()| function in \verb|extension.ts| is called. It creates a \verb|gutterIcon| to be displayed beside function signatures and reads the configuration file using the\verb|readConfig()| helper function from \verb|input.ts|. It also registers commands: \verb|continue|, \verb|validate|, and \verb|exit|, which will be used in later functions. Next, we subscribe to event listeners for changes in the editor (the pane), document, and cursor. When the active text editor changes or a text document belonging to the workspace is modified, the \verb|checkFns()| function from our \verb|utils.ts| is called. It finds all function signatures in the editor that match the specified function name from the configuration and creates a decorations object for each.

When the \verb|onDidChangeTextEditorSelection| event is triggered, and the user's cursor is found to be on a line of a supported function, an \verb|informationMessage| is displayed prompting the user to begin the learning process. If started, the \verb|sandbox()| function in \verb|utils.ts| is called. It creates a temporary Python file by extracting the function and imports from the active document (using \verb|getImports()| and \verb|getFnRange())| and generating an entry point (the \verb|genMain()| function). We await its completion to hide the current editor, then run \verb|dim()|, \verb|focus()|, and \verb|startDebugger()| from our \verb|utils.ts|.

If the user chooses to continue to the next line, the compass.continue command is executed. It calls the focus function again with the next line number and updated configuration. The startDebugger function in  \verb|utils.ts| is called to set a breakpoint on the line after the focused line and starts the debugger. If the user reaches the end of the learning process, a CodeLens is registered bound to command \verb|compass.validate|, (implementation incomplete). Every CodeLens is registered with the compass.exit command, which returns to the last editor \verb|origEditor|, calls \verb|exitDebugger|, and disposes of the temp file.

\begin{figure}
    \centering
    \includegraphics[width=8cm]{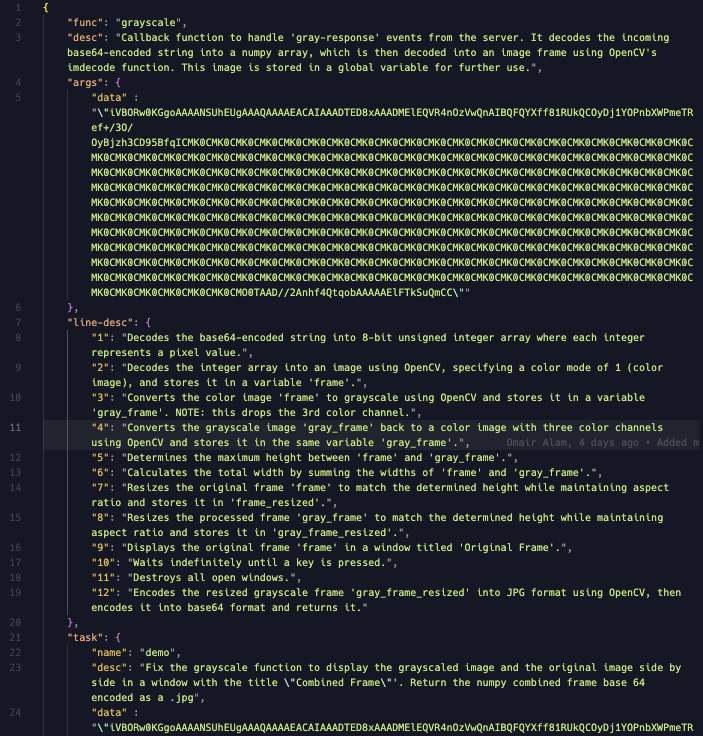}
    \caption{Config file for our codebase}
    \label{fig:config}
\end{figure}

We define a config file that houses all the information needed to run our extension on particular codebase. This includes the function name, corresponding documentations, and tasks (as shown in Figure \ref{fig:config}). For each task, we even store the serialized inputs and outputs that we can use when validating tasks with the user input.

\section{Technical Challenges}

\subsection{Challenge 1: Parsing Python}

In order to make Python functions interactable and build a 'sandbox' around a function, we needed to a way to parse the active python file and extract function signatures, the body, and imports. Unfortunately, VSCode's Python extension does not expose its AST, and there are no mature/reliable Node.js libraries for the task either.

But because VSCode extensions are based on subscribing to watchers which are fairly broad, (\verb|onDidChangeTextDocument|, \\ \verb|onDidChangeTextEditorSelection|), states need to be built on each change anyway. The simplest and fastest approach, then, was to perform a regex pattern match to find imports and function signatures. Given the starting line of a function signature, we easily find where its body ends by skipping empty lines and finding the first line where the indentation is less than the starting line.

\subsection{Challenge 2: Integrating with Python Debugger}

Our goal was to implement a 'sandbox' mode that allows users to engage directly with our codebase, aiming to achieve two main objectives:

\begin{enumerate}
    \item Enable users to step through functions line-by-line, displaying descriptions of each line's purpose adjacent to the code.
    \item Allow users to complete tasks within the sandbox to deepen their understanding of the functions.
\end{enumerate}

To address the first objective, we integrated line-by-line descriptions with two existing Visual Studio Code (VSCode) functionalities: the Python Debugger (PDB), which facilitates step-by-step code navigation, and the VSCode Debugger Graphical User Interface (Debug GUI), which displays variable states at various stages.

Although we found an API to initiate PDB programmatically, there was no API available to simultaneously initialize the Debug GUI. Starting PDB alone would have required users to check variable states via the Terminal/Console, which would detract from the user experience. Developing our own GUI was not a feasible option either, as it would unnecessarily duplicate many of the features already provided by the existing Debug GUI.

Without a suitable API, we chose to utilize VSCode's keybindings to activate the Debug GUI and navigate the debugger line-by-line. Initially, we considered manually triggering specific keystrokes, such as F5 to start the Debug GUI, but we later found that VSCode had named commands for these actions. Thus, we opted to use these named commands instead, significantly streamlining our workflow:

\begin{enumerate}
    \item When a user starts the sandbox tool, we issue the VSCode command to launch the Debug GUI.
    \item When a user clicks 'Continue' in our tool, we issue another command to advance the debugger to the next line.
\end{enumerate}

This approach ensures that the line descriptions in our tool align perfectly with the debugger's current line, providing a seamless user experience. We also included functionality in the tool integration with the Debug GUI that allows the user to step through and step over lines, further enhancing interaction capabilities

\section{Discussion}

Following up on our initial need-finding study, which underscored significant challenges that software engineers face while debugging unfamiliar codebases, we pinpointed a prevalent issue: participants struggled to effectively use relevant documentation when navigating through codebases. Commonly, they turned to split-screen setups or frequent web searches, leading to substantial context switching and fragmented comprehension. CodeCompass addresses these issues by integrating documentation directly within the Integrated Development Environment (IDE) and providing context-aware suggestions and visualizations. This integration allows developers to access necessary information without exiting their coding environment, fostering focused work and significantly reducing disruptive context switching.

To evaluate the effectiveness of CodeCompass, we conducted a formative study using a setup similar to that of the initial need-finding study, but with a subset of participants who were specifically recruited for this phase of evaluation. This approach ensured that individuals who were already familiar with the task and its inherent challenges could provide informed feedback on the tool's impact. Participants were given the same codebase and assigned identical bug-fixing tasks as those in the initial study. However, in this iteration, they were introduced to CodeCompass and then observed as they used the tool to complete their tasks.

During the formative study, researchers observed participants' interactions with CodeCompass, noting their navigation patterns, information-seeking behaviors, and any challenges they encountered. This comprehensive observation helped us gather critical insights into how CodeCompass enhances the debugging experience and supports developers in navigating and comprehending code more effectively.

The completion of the formative study revealed promising results with respect to the effectiveness of CodeCompass in enabling users to identify code comprehension challenges. Observations of participants' interactions with the tool were able to provide valuable insight into the impact it made on their debugging process in a new codebase. 

Both participants in the formative study were able to identify the root cause of the bug by recognizing that the function wasn't working as intended. However, their approaches to solving the problem differed demonstrating the flexibility of CodeCompass to serve as a tool for their choice of learning. P1 leveraged the tool's walkthrough feature to get a better understanding of the functions' logic and identified points of failure. The step by step explanations and visualizations provided by CodeCompass served in P1's eyes as “better comments”. P2 decided to utilize the in line descriptions along with the input/output case from the sandbox to confirm his hypothesis. This let him carry out a targeted search for debugging the exact issue faced in the code. These different approaches demonstrate the tool, CodeCompass, can allow developers to carry out their preferred learning style when tasked with completing a new codebase.

A key observation from the formative study is that we saw a reduction in the context switching among participants using CodeCompass. By providing an in-line documentation debugger and a sandbox environment within the IDE, the tool enables developers to stay focused on the task at hand and avoid disruptive transitions between different applications or having to split screen.
	
The formative study on the CodeCompass tool displayed that integrating documentation into the debugger improves the developer's code comprehension and efficiency. Secondly, it releases a reduction of context switching which results in increased focus and enables a more seamless debugging experience. The tool's ability to function with a different learning style enables its use for a variety of users. 

Reflecting on the process of integrating documentation and learning tools into software development environments, several key learnings emerge that would inform how we might approach similar projects differently in the future:

\begin{enumerate}

    \item Deep Understanding of User Environment: Gaining a deeper understanding of the actual working conditions and needs of software developers through comprehensive need-finding studies is crucial. Knowing what we do now, investing more time upfront to understand user workflows would guide the design to better fit into existing environments and meet user needs more precisely.
    
    \item Seamless Integration: The importance of integrating enhancements directly within tools that developers are already comfortable with became clear. In future projects, focusing on augmenting existing platforms rather than introducing separate tools would minimize disruption and enhance adoption.
    
    \item Iterative Development and Feedback: The value of an iterative design and testing process was reinforced. Moving forward, planning for more frequent iterations and incorporating user feedback at each stage would help in refining the tool more effectively and addressing practical issues earlier.
    
    \item Real-world Evaluation: The real-world performance of the tool provided critical insights that were not apparent in controlled settings. In future projects, conducting more extensive real-world evaluations would be key to understanding the long-term impact and practical viability of the tools developed.
  
    \item Thorough Documentation and Knowledge Sharing: The importance of documenting the process, decisions, and outcomes thoroughly became evident. Better documentation practices would facilitate knowledge transfer and provide a valuable resource for ongoing development and future enhancements.
\end{enumerate}

\section{Future Study}

When developing tasks for our function simulator, one generalizable approach could involve utilizing Large Language Models (LLMs) with contexts from the code repository and associated documentation. However, evidence indicates that LLMs may not effectively generate meaningful learning tasks. In light of this, as we progress towards a deployable version of our tool, we plan to thoroughly review literature concerning what defines effective learning tasks and best pedagogical practices. Our goal is to tailor our learning goals based on these educational standards. Moreover, we intend to scrape and analyze data from Stack Overflow, which will help us identify prevalent bugs and extract insights from the ways people frame their questions and the responses that receive significant community approval.

Expanding on the vision for enhancing our function simulator and addressing the challenges associated with utilizing LLMs in generating learning tasks, we recognize the need to explore several innovative research directions. While LLMs present a promising approach for auto-generating context-aware tasks based on the documentation and code repositories, their current limitations in generating truly meaningful and educational tasks necessitate a broader, multi-faceted research strategy.

\begin{enumerate}
    \item Advanced Machine Learning Techniques: Beyond basic LLMs, we can investigate the integration of more sophisticated machine learning models that are specifically tailored for educational content creation in software development. Techniques such as reinforcement learning could be employed to refine the model's ability to generate tasks that not only align with the user's current understanding but also challenge their skills progressively.

    \item Interactive Learning Environments: Enhancing the interactivity of the function simulator by incorporating real-time feedback mechanisms could significantly improve the learning experience. This could involve the development of an adaptive learning system that adjusts the difficulty of tasks based on the user's performance and engagement levels.

    \item Integration with Development Workflows: Research into how such tools can be seamlessly integrated into typical software development workflows without disrupting existing practices. This could involve developing plugins or extensions for popular IDEs other than VS Code, considering their API limitations. It could also involve creating standalone applications that sync with development environments through APIs that allow more flexible integration options.

    \item Pedagogical Validation: Conduct empirical studies to validate the educational effectiveness of the generated tasks. This involves setting up controlled experiments with real users to measure improvements in their understanding and retention of programming concepts, as well as their ability to apply what they've learned in practical scenarios.

    \item Community-Driven Development Models: Explore the potential of crowd-sourced models where experienced developers can contribute to task generation and validation. This could help ensure that the tasks are not only technically accurate but also pedagogically effective. Leveraging platforms like GitHub could facilitate community involvement in refining and enhancing the learning tasks.

    \item Utilization of Domain-Specific Languages (DSLs): Investigate the use of DSLs within LLMs to improve the specificity and relevance of generated tasks. By training models on domain-specific data, tasks generated could be more aligned with real-world programming scenarios, providing more practical learning experiences.

    \item Exploring Other Data Sources: Beyond Stack Overflow, consider analyzing discussion forums, technical blogs, and other coding communities like Reddit's r/programming or Hacker News to gather a broader range of insights and common challenges faced by developers. This could enrich the dataset used for training our models, leading to more diverse and comprehensive task generation.
    
    \item Ethical and Bias Considerations: As part of our research, it will be essential to consider the ethical implications and potential biases in machine-generated educational content. Ensuring that the learning tasks are inclusive and equitable will be crucial, necessitating regular audits and updates to the training data and model algorithms.
    
    \item Longitudinal Studies on Learning Impact: Initiate long-term studies to track the progress and outcomes of learners using the tool. These studies would help in understanding how effectively the tool helps bridge knowledge gaps over time and how it impacts the overall development process of the learners.
\end{enumerate}

By broadening the scope of our research to include these areas, we aim to significantly enhance the capability of our function simulator to provide not only relevant and challenging tasks but also to do so in a manner that is pedagogically sound and deeply integrated with the software development life-cycle.
\begin{acks}
To Prof. Sarah Chasins whose elevator insight and guidance was instrumental in the formulation of our study.
\end{acks}

\bibliographystyle{ACM-Reference-Format}
\bibliography{sample-base}


\end{document}